\documentclass[
reprint,
superscriptaddress,
citeautoscript,
showpacs,
nofootinbib,
nobibnotes,
amsmath,amssymb,
aps,
floats,
floatfix,
]{revtex4-1}

\usepackage{graphicx}
\usepackage{dcolumn}
\usepackage{bm}

\begin{document}

\renewcommand{\thefootnote}{\alph{footnote}}

\title{Micromagnetic study of a feasibility of the magnetic anisotropy engineering\\ in nano-structured epitaxial films of (III,Mn)V ferromagnetic semiconductors}

\author{K. Dziatkowski}
 \email[Corresponding author: ]{konrad.dziatkowski@fuw.edu.pl}
\author{A. Twardowski}
\affiliation{Faculty of Physics, University of Warsaw, 00-681 Warsaw, Poland}

\date{\today}

\begin{abstract}
The attainability of modification of the apparent magnetic anisotropy in (III,Mn)V ferromagnetic semiconductors is probed by means of the finite-elements-based modelling. The most representative case of (Ga,Mn)As and its in-plane uniaxial anisotropy is investigated. The hysteresis loops of the continuous films of a ferromagnetic semiconductor as well as films structured with the elliptic anti-dots are modelled for various eccentricity, orientation, and separation of the anti-dots. The effect of anti-dots on the magnetic anisotropy is confirmed but overall is found to be very weak. The subsequent modelling for (Ga,Mn)As film with the elliptic dots comprising of metallic NiFe shows much stronger effect, revealing switching of the magnetic moment in the ferromagnetic semiconductor governed by the switching behavior of the metallic inclusions.
\end{abstract}

\pacs{75.50.Pp, 75.30.Gw, 75.75.-c, 75.78.Cd}

\maketitle

\section{\label{sec:intro}INTRODUCTION}
It has been almost three decades since the seminal works of Munekata \emph{et al.} \cite{Munekata} that started the remarkable series of the experimental and theoretical studies of the ferromagnetic epitaxial layers of dilute magnetic semiconductors (DMS). Among them the family of ferromagnetic (III,Mn)V alloys is arguably the one studied most extensively, despite the fact that Curie temperature of paramagnet-to-ferromagnet transition revealed in these materials is still well below $300$ K, a factor hindering the postulated suitability of (III,Mn)V semiconductors for future electronic devices \cite{Akinaga,Dietl}. Nevertheless, these materials, and (Ga,Mn)As in particular, remain a testbed for magnetic and transport phenomena being prerequisites of DMS-based approach to electronics, including but not limiting to the engineering of magnetic anisotropy \cite{Liu}.

The native magnetic anisotropies of DMS have been studied extensively and quite successfully, to mention only the long debate about the origin of the uniaxial in-plane component of the magnetic anisotropy in (Ga,Mn)As \cite{Birowska}. In pursuit of the controllable engineering of magnetic anisotropy in (III,Mn)V the various studies reaching beyond the single, continuous film paradigm were conducted. Among them one may find numerous examples of works making use of strain or proximity effects \cite{King,Dziatkowski2006}, but those trying the magnonic crystal approach \cite{Lenk} are remarkably sparse \cite{Dziatkowski2016}. This paper aims on contributing in filling the gap. It investigates, by means of micromagnetic modelling, a feasibility of the controllable modification of the in-plane magnetic anisotropy in ferromagnetic (III,Mn)V semiconductor. The anisotropy engineering is approached with anti-dot structuring as well as with embedding dots of a soft ferromagnet. It is one of the first attempts to "transplant" the idea of magnonic structures to DMS from the domain of metallic ferromagnets, where it has been very fruitful \cite{Kruglyak,Neusser}. It should be enough to mention that such nano- and micro-engineered metallic systems with periodic modification of magnetic properties revealed, for example, magnetic anisotropies of exotic, non-native symmetries, a feature of the particular interest in the context of this paper \cite{Crowburn,Tacchi}.

\section{METHODS}
For the purpose of this study a $50$-nm-thick layer of (III,Mn)V semiconductor was modeled, with the saturation magnetization $M_{\mathrm{sat}} = 51.4\cdot10^3$ A/m. The chosen value of $M_{\mathrm{sat}}$ corresponds to (Ga,Mn)As with $5$\% of manganese, under assumption that all magnetic ions substitute gallium and contribute to the saturation magnetization. Whereas it is well known experimental fact that average magnetic moment per Mn in (Ga,Mn)As is less than 5$\mu_{\mathrm{B}}$, such simplification is easily accepted when one remembers that $M_{\mathrm{sat}}$ is only a scaling factor in micromagnetic modeling and does not invalidate the method of tailoring of the magnetic anisotropy as pursued here. Furthermore, the above value of the saturation magnetization, while significant, is not uncommon in the experimental studies of (III,Mn)V (see for example Ref. \onlinecite{HailorWang} and references within). Thus it should be regarded as one of the probable (i.e. not distinctive) values for modeled (III,Mn)V materials. The same rule applies to the anisotropy and exchange parameters discussed below. 

The magnetic anisotropy model was assumed according to the acknowledged work of Farle \cite{Farle} featuring both cubic and uniaxial components:
\begin{eqnarray}
F & = & -\frac{M}{2}\cdot\left[H_{2\perp}\cos^2\theta_M+\frac{1}{2}H_{4\perp}\cos^4\theta_M+\right.\nonumber\\
& & +\frac{1}{8}H_{4\parallel}\left(3+\cos(4\phi_M)\right)\sin^4\theta_M+\\
& & +\left.\frac{1}{2}H_{2\parallel}\left(1-\sin(2\phi_M)\right)\sin^2\theta_M\right];\nonumber
\label{eq:mae}
\end{eqnarray}
$\theta_M$ and $\phi_M$ denote, respectively, the polar and azimuthal angle of the magnetization vector. The assumed anisotropy fields, $H_{i}$, are presented in Table \ref{tab:params}.
\begin{table}[b]
\caption{\label{tab:params}The anisotropy fields, $H_{i}$, of magnetic anisotropy energy $F$ described by Eq. \ref{eq:mae}. The corresponding anisotropy constants can be easily derived from $K_{i} = H_{i}M_{\mathrm{sat}}/2$.}
\begin{ruledtabular}
\begin{tabular}{ccccc}
\vspace{0.1cm}
&$H_{2\parallel}$ &$H_{2\perp}$ &$H_{4\parallel}$ &$H_{4\perp}$\\
\hline
\vspace{0.1cm}
CGS & 0 G & -2500 G & 1000 G & 300 G\\
SI & 0 kA/m & -199 kA/m & 80 kA/m & 24 kA/m\\
\end{tabular}
\end{ruledtabular}
\end{table}
One should note that in-plane uniaxial anisotropy component $H_{2\parallel}$ --- frequently observed in the epitaxial layers of dilute ferromagnetic semiconductors --- was set equal to zero, since demonstration of the engineered uniaxial anisotropy is the main goal of this paper. Secondly, the out-of-plane cubic anisotropy field, $H_{4\perp}$, is not relevant when only in-plane effects are discussed, nevertheless it is shown in the table and included in calculations for the sake of completeness of the model.

The last parameter to be defined in the model should be either exchange length, $\lambda_{\mathrm{ex}}$, or exchange constant, $A$,\footnote{One should be aware that some authors use the name "exchange stiffness" to describe the constant defined here. We follow the prevailing nomenclature where the exchange constant, $A$, differs from the exchange stiffness, $D$, as follows: $A = M_{\mathrm{sat}}D/2$; see for example Ref. \onlinecite{HoffmannPhD}.} which are related to each other via the following formula:
\begin{equation}
A = \frac{\mu_0}{2}M_{\mathrm{sat}}^2\lambda_{\mathrm{ex}}^2.
\label{eq:exchange}
\end{equation}
The available data for exchange constant in (III,Mn)V vary greatly depending on the material in question ((Ga,Mn)As, (Ga,Mn)(As,P), etc.), its properties (manganese content, saturation magnetization, hole concentration, etc.), and derivation method (theoretical estimates, experimental spin wave dispersion, etc.). In this study the exchange constant $A = 0.37$ pJ/m was used, which may be regarded as rather high for (Ga,Mn)As, for which the values in the range of $0.20-0.24$ pJ/m were typically reported in both experimental and theoretical research \cite{HoffmannPhD,Jungwirth,WerpachowskaPhD}. However, for (Ga,Mn)(As,P) even larger values of the exchange constant were seen \cite{Haghgoo}. Moreover this value of $A$ corresponds to the exchange length $\lambda_{\mathrm{ex}}$ of about $15$ nm, comparable to the values reported before, for example by Hoffmann \cite{HoffmannPhD}.

From the viewpoint of micromagnetic modeling, the actual value of the exchange length is not so important as the relation of this quantity to the parameters of finite elements mesh on which the modeling is performed. First of all, $\lambda_{\mathrm{ex}}$ should be significantly larger than the distance between the nearest neighbours of the mesh, $d_{\mathrm{NN}}$. For the results presented below, this condition was fulfilled by using meshes with $d_{\mathrm{NN}}=3.0$ nm, which is one-fifth of $\lambda_{\mathrm{ex}}$. On the other hand, it would be beneficial if the characteristic sizes of features structured in a modeled film were larger than but comparable to the exchange length, since then the effects of discontinuity could be most pronounced. For that reason the lateral sizes of the anti-dots were kept between $\lambda_{\mathrm{ex}}$ and $4\lambda_{\mathrm{ex}}$.

\begin{figure}[!h]
\includegraphics[width=0.35\textwidth]{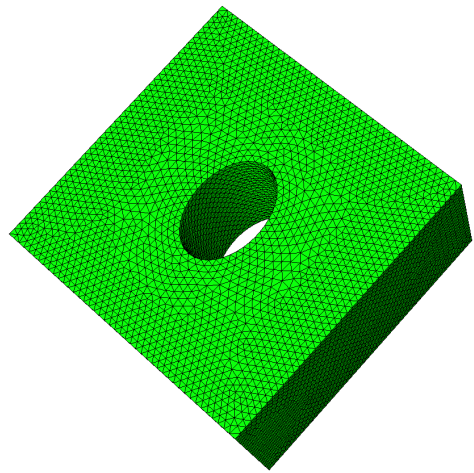}
\caption{\label{fig:mesh}Finite elements mesh with an anti-dot of axes equal $36$ and $25$ nm. The cuboid edges are along $[100]$ (width), $[010]$ (depth), and $[001]$ (height) crystallographic directions.}
\end{figure}

In order to limit the number of varying parameters, the anti-dots were modeled as $50$-nm-deep (i.e. all-through) ellipses of a constant area equaling $\pi15^2$ nm$^2$. With such constraint, the "missing" volume of the film was always the same, whereas the eccentricity of the anti-dot --- changing between $0$ and $0.97$ --- was the independent variable influencing the apparent magnetic anisotropy. The axes of the ellipses were kept along $[100]$ (major) and $[010]$ (minor) in-plane crystallographic directions, with $[001]$ being the normal ("growth") axis. In Fig. \ref{fig:mesh}, a mesh with an anti-dot of intermediate value of eccentricity is presented.
The cuboid surrounding the anti-dot and modeling the remaining (III,Mn)V material in each case had the width and depth of $100$ nm and the height of $50$~nm. Such cuboid was repeated 9 times both along $[100]$ and $[010]$ directions, providing so-called macrogeometry, i.e. quasi periodic boundary conditions \cite{Fangohr}. Initially, the size of the macrogemetry was varied and it was confirmed that $9$-by-$9$ lattice was enough to ensure stability of the results.

Dynamics of the magnetization $\mathbf{M}$ in the system described above is governed by Landau-Lifshitz-Gilbert equation ($\gamma$ - the gyromagnetic ratio, $\alpha$ - the damping parameter)
\begin{eqnarray}
\frac{\partial\mathbf{M}}{\partial t} = -\gamma\mathbf{M}\times\mathbf{H} + \alpha\frac{\mathbf{M}}{M}\times\frac{\partial\mathbf{M}}{\partial t},
\label{eq:LLG}
\end{eqnarray}
where $\mathbf{H}$ is the effective magnetic field containing also contributions other than the external field, e.g. from the crystalline anisotropy. With use of NMag micromagnetic simulator Eq. (\ref{eq:LLG}) was explicitly integrated in the consecutive time steps at each node of the finite element mesh \cite{Nmag}. Since main interest of this study lays in the analysis of hysteresis loops, i.e. in the equilibrium direction of the magnetization vector, the modeled system was artificially over-damped by setting the damping parameter $\alpha=0.3$, which is about $10$ times larger than values reported in the literature for (III,Mn)V (see for example Ref. \onlinecite{Nemec}). In such a way the simulations were speed up significantly with no harm to the quality of static results.

\section{RESULTS AND DISCUSSION}
\begin{figure}[!h]
\includegraphics[width=0.45\textwidth]{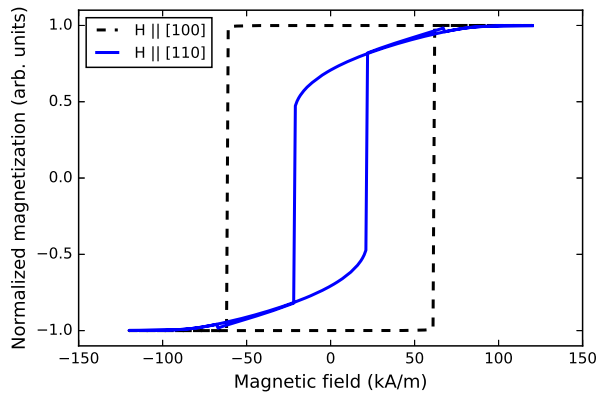}
\caption{\label{fig:solid}Hysteresis loops --- normalized to the saturation magnetization --- modeled for the external magnetic field $\mathbf{H}$ applied along two different crystallographic directions of an unstructured film.}
\end{figure}

Modeling of the system of interest started with an unstructured film, which normalized hysteresis loops are presented in Fig. \ref{fig:solid}.
One can easily notice that $[100]$ is the easy axis of the magnetization featuring distinctly rectangular hysteresis loop and the coercive field of about $61$ kA/m. On the other hand, the $[110]$ direction is the harder one, with the coercive field of only about $21$ kA/m and the hysteresis loop characterized by long saturation tails. It was confirmed, that $[010]$ and $[1\overline{1}0]$ directions are equivalent to $[100]$ and $[110]$, respectively, since in-plane uniaxial anisotropy field $H_{2\parallel}$ was set to zero.

\begin{figure}[!h]
\includegraphics[width=0.45\textwidth]{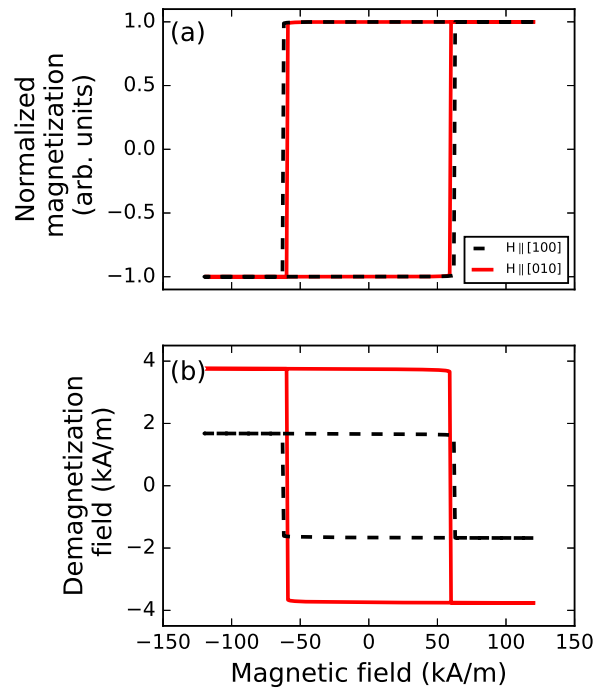}
\caption{\label{fig:hole30}(a) Hysteresis loops of the film with an elliptical anti-dot which major and minor axes were $60$ and $15$ nm in length and oriented along $[100]$ and $[010]$, respectively. (b)~Corresponding demagnetization field averaged over the entire modeled film.}
\end{figure}

Then, the hysteresis loops analogous to those presented in Fig. \ref{fig:solid} were simulated for the film with circular anti-dot of $15$ nm radius. The loops appeared to be overlapping with those generated for the unstructured (continuous) film, indicating there is no significant effect of the circular anti-dot when averaged over the sample (obviously, one may expect some local effect of discontinuity at the edge of an anti-dot). In the next steps, the films with anti-dots of elliptical cross-section were investigated, and Fig. \ref{fig:hole30}(a) presents the hysteresis loops for one of such films.
One should remind that the major and minor axes of the ellipse were along $[100]$ and $[010]$ crystallographic directions, respectively. The loops simulated for the external magnetic field applied along these directions differ in width by about $6$ kA/m. Whereas the width difference is unmistakable, the observed value is quite small and only few times larger than the simulation step ($1$ kA/m) what makes it difficult for further analyses. Much more convenient is investigation of the demagnetization field (averaged over the film), which is presented in Fig. \ref{fig:hole30}(b). Now, when looking at both panels of Fig. \ref{fig:hole30}, one can conclude that anti-dot structuring results in the hardening of the direction along the minor axis (here $[010]$), which is manifested by the narrower hysteresis loop and can be attributed to the larger (as compared to $[100]$ direction) demagnetization field countering the applied external magnetic field. This effect can be regarded as the structure-induced in-plane uniaxial anisotropy (let us remind that no intrinsic in-plane uniaxial anisotropy was assumed in this study).

\begin{figure}[!h]
\includegraphics[width=0.45\textwidth]{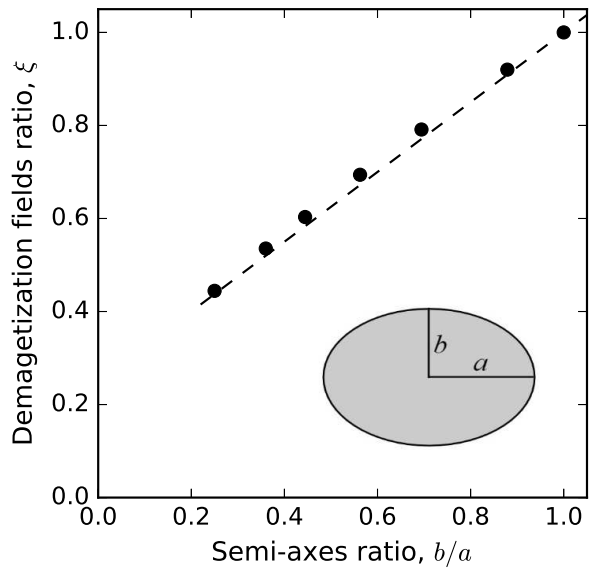}
\caption{\label{fig:demag_ratios}Ratio of demagnetization fields at $H=0$ for films with elliptic anti-dots of various ratio of the semi-axes. Dashed line is just a guide for the eye.}
\end{figure}

With demagnetization field data in place, one can investigate quantitatively how the shape of anti-dots affects the structuring-based contribution to the in-plane magnetic anisotropy. It is know from Ref. \onlinecite{Beleggia}, that demagnetization field of the infinite elliptic cylinder magnetized uniformly in the direction perpendicular to its side surface depends on the ratio of the cylinder's major semi-axes, $b/a$. Whereas this paper deals with a "missing" portion of the film in the form of elliptic cylinder of finite length, the semi-axes ratio, or compression factor, of an anti-dot is expected to be a proper independent variable for shape-effect study. Let us define $\xi$ as the ratio of film-averaged demagnetization fields registered along major ($[100]$) and minor ($[010]$) semi-axes at $H=0$:
\begin{eqnarray}
\xi=H_{demag}^{[100]}/H_{demag}^{[010]}.
\label{eq:Xi}
\end{eqnarray}
Then Fig. \ref{fig:demag_ratios} shows how $\xi$ depends on the anti-dot compression factor.
It is evident that more eccentric anti-dots induce stronger uniaxial anisotropy between $[100]$ and $[010]$ directions, although it is hard to argue for a meaning of the almost linear dependence of $\xi(b/a)$. Data in the plot stop short of $b/a=0.2$ for two reasons. Smaller compression factors would yield ellipses which major axes are close to the lateral size of the modelled cuboid and which sharp endings would have to be surrounded by the very dense finite elements mesh. For $b/a<0.2$ the apparent linear dependence of $\xi$ should change towards the plot's origin $(0,0)$, since in the limit of $b/a=0$ the ellipses take form of trenches cutting the film into stripes which width is equal to the width of the cuboid and length is "infinite" under periodic boundary conditions. In such a case the nominator of $\xi$ equals zero while the denominator remains finite.

\begin{figure}[!h]
\includegraphics[width=0.45\textwidth]{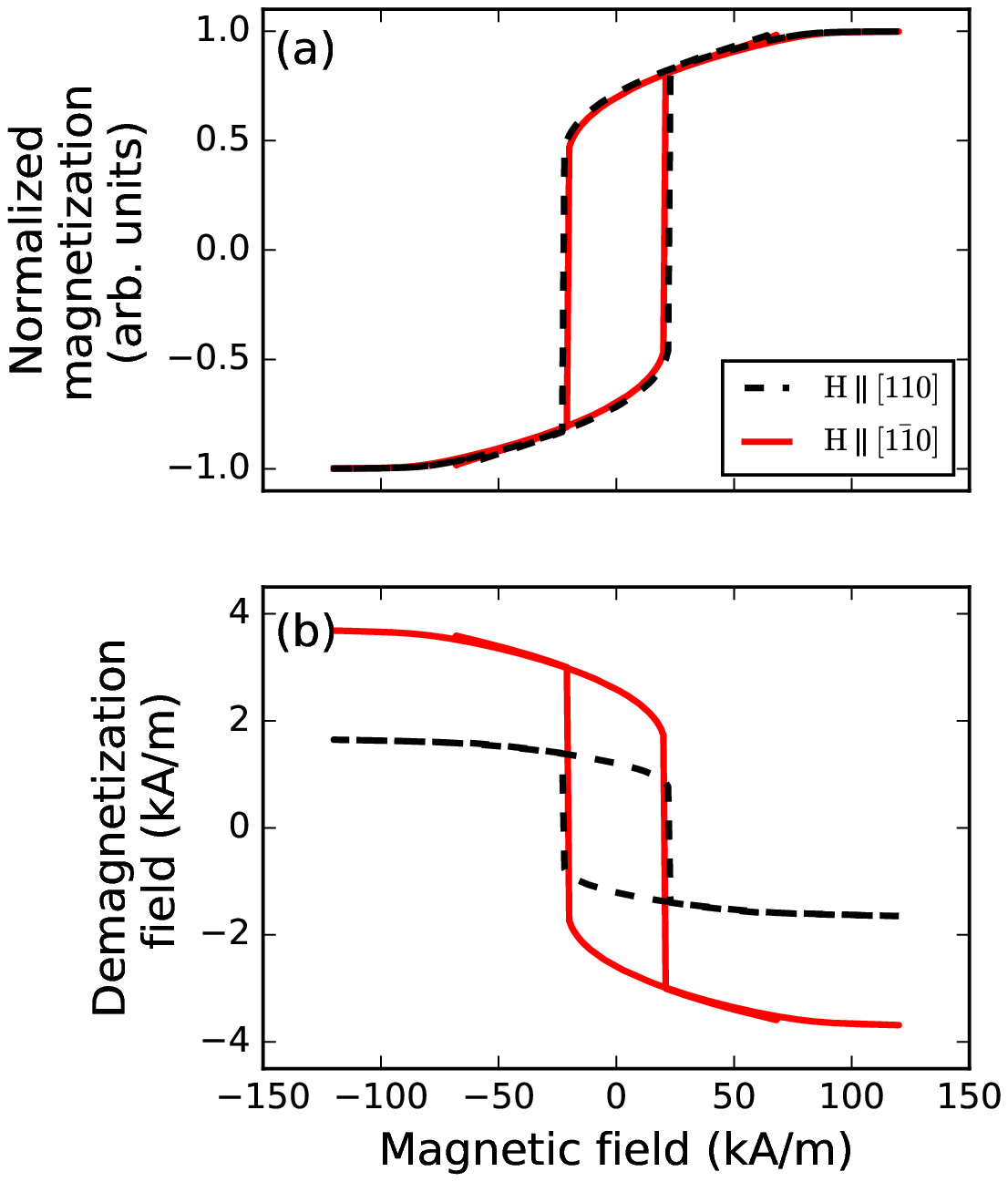}
\caption{\label{fig:diagonal30}(a) Hysteresis loops of the film with an elliptical anti-dot which major and minor axes were $60$ and $15$ nm in length and oriented along $[110]$ and $[1\overline{1}0]$, respectively. (b)~Corresponding demagnetization field averaged over the entire modeled film.}
\end{figure}

In the canonical dilute ferromagnetic semiconductor (Ga,Mn)As the intrinsic in-plane uniaxial anisotropy is usually observed between $[110]$ and $[1\overline{1}0]$ crystallographic directions \cite{Welp} (although other types had been also reported; see for example Ref. \onlinecite{Pappert}). Thus for the sake of completeness the analogous modelling was performed for the film with the elliptic anti-dots which major axes were oriented along $[110]$. The results are presented in Fig. \ref{fig:diagonal30}.
The structuring-induced anisotropy of both the hysteresis loop and demagnetization field was observed but the effect of anti-dots on the former was of very limited magnitude, similarly to Fig. \ref{fig:hole30}. The hysteresis loop difference visible in the upper panels of Figs. \ref{fig:hole30} and \ref{fig:diagonal30} ($6$ kA/m or $75$ Oe), is practically at the discrimination threshold achievable in a typical magnetometer. Thus in the following paragraphs the ways to enhance the effect of structuring will be discussed.

The most simple approach would be to increase the density of structuring. The results presented above were collected for ferromagnetic films from which about $7$\% of volume was removed by anti-dots. Indeed, when one increases the density of structuring to about $14$\%, the difference in width of the hysteresis loops recorded along the major and minor axes rises. For $60$-by-$15$ nm anti-dot it reaches about $16$-$68$ kA/m (or $200$-$850$ Oe), as compared to $6$ kA/m reported above. The actual value depends on whether the anti-dot spacing is reduced by the same factor along $[100]$ and $[010]$. But such a straightforward approach is unappealing. The method for engineering of the magnetic anisotropy discussed in this paper is only useful when the amount of anti-dots is limited, so after the structuring one still ends with a film with only some modifications.

\begin{figure}[!h]
\includegraphics[width=0.45\textwidth]{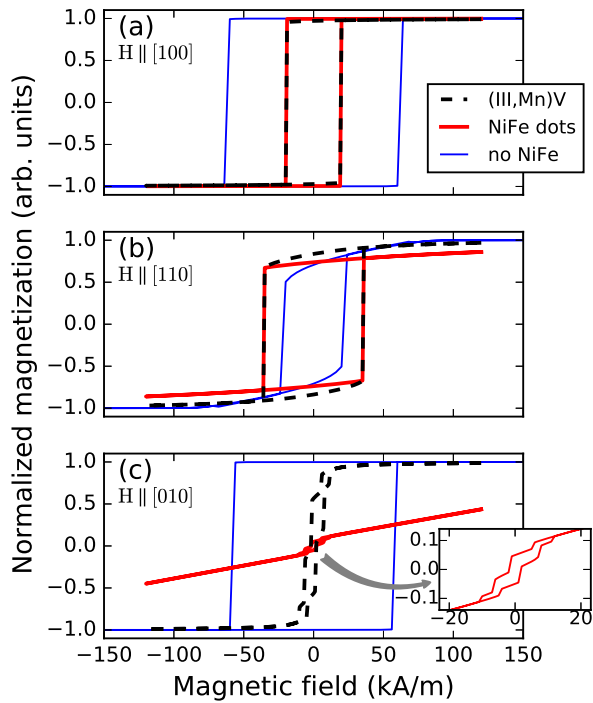}
\caption{\label{fig:dotsPy}Hysteresis loops collected for (III,Mn)V film with elliptical NiFe dots which major and minor axes were $50$ and $18$ nm in length and oriented along $[100]$ and $[010]$, respectively. The loops, corresponding to (III,Mn)V (black) and NiFe (red) were presented for the external magnetic field applied along (a) [100], (b) [110], and (c) [010] crystallographic directions. The blue loops, corresponding to (III,Mn)V with the analogous $50$-by-$18$ anti-dots, were added for comparison.}
\end{figure}

The alternative way is to go beyond the anti-dots, e.g. to fill them with a soft magnetic material like Ni$_{0.8}$Fe$_{0.2}$. In the next steps it was assumed that such metallic inclusions show no instrinsic, magnetocrystalline anisotropy and are characterized by the magnetization $M^{\mathrm{NiFe}}_{\mathrm{sat}} = 8\cdot10^5$~A/m and the exchange constant $A^{\mathrm{NiFe}} = 13$~pJ/m \cite{Ingvarsson,Pan}. Figure \ref{fig:dotsPy} presents the hysteresis data for (III,Mn)V film with the elliptic $50$-by-$18$ nm NiFe dots, as compared to the analogous data for the film with the anti-dots of identical shape and orientation.
Note that these loops depict the normalized magnetization \emph{separately} for (III,Mn)V and NiFe, i.e. it is not the magnetization averaged over the entire sample consisting of both materials. When looking at (III,Mn)V data, the difference between $\mathbf{H}\parallel[100]$ and $\mathbf{H}\parallel[010]$ directions --- which correspond to the major and minor axes of NiFe dots, respectively --- is significant and undoubtedly visible. For $\mathbf{H}\parallel[100]$ (top panel of Fig. \ref{fig:dotsPy}) the hysteresis loop is sharp, square-like, although much narrower than the one observed for the film without NiFe inclusion. Apparently, the $[100]$ direction remains as the easy-axis of (III,Mn)V but it seems that the switch of hysteresis loop, i.e. the coercive field, is governed by the switch of the magnetic moment in NiFe inclusions. This claim holds true also for $\mathbf{H}\parallel[110]$ (middle panel of Fig. \ref{fig:dotsPy}), where again the coercive fields of NiFe dots and (III,Mn)V matrix (in which these dots are embedded) coincide, whereas both are different than the coercive field of (III,Mn)V film with anti-dots. One should note that at $H\approx0$ the normalized magnetization of (III,Mn)V matrix is slightly higher than the corresponding value for (III,Mn)V with anti-dots. It is a manifestation of NiFe magnetic moment bending (III,Mn)V magnetic moment from $[010]$ (which is the nearest, local minimum of energy for $\mathbf{M}$ in a film with anti-dots) towards $[100]$ (which is the major axis of the dots), thus increasing the projection of (III,Mn)V magnetic moment on the direction of $\mathbf{H}$.

\begin{figure}[!h]
\includegraphics[width=0.45\textwidth]{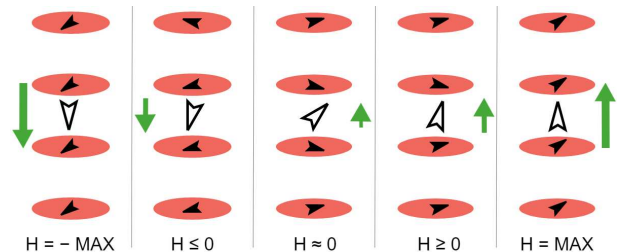}
\caption{\label{fig:dotsPyExplained}Explanation of the features observed in the hysteresis loops collected in Fig. \ref{fig:dotsPy}(c): green arrows --- the external magnetic field; solid black arrows --- the magnetic moment of NiFe; empty black arrows --- the magnetic moment of (III,Mn)V. See text for detailed discussion.}
\end{figure}

The interplay between the magnetic moments of NiFe dots and (III,Mn)V matrix is even more evident in the complicated shape of the hysteresis loops depicted in the bottom panel of Fig. \ref{fig:dotsPy}, i.e. for the external magnetic field applied along the minor axis of the dots. The cartoon of Fig. \ref{fig:dotsPyExplained} will be helpful in this analysis.
The relative magnitude of the external magnetic field $\mathbf{H}\parallel[010]$ is depicted by green arrows and is swept from high to low (right-to-left). The solid black arrows show the orientation of the magnetic moment of NiFe dots whereas the empty black arrows shows the orientation of the magnetic moment of (III,Mn)V material surrounding these dots. At the highest magnetic fields, (III,Mn)V is saturated along $\mathbf{H}$ whereas the magnetic moment of NiFe assumes an intermediate orientation between the direction of $\mathbf{H}$ and the major axis of the dots. When the external magnetic field is changed to smaller but still positive values --- around the right-most step visible in the down-swept part of the hysteresis loops in Fig. \ref{fig:dotsPy}(c) --- some of NiFe dots switch orientation of their magnetic moment. At the same time the magnetic moment of (III,Mn)V matrix is bend out of direction of $\mathbf{H}$ toward the major axis of the dots. This process continues in a step-like manner with further decrease of $H$. At some negative fields it is more favorable for (III,Mn)V to switch its magnetic moment to be more aligned with $\mathbf{H}$ and with the projection of NiFe moment on $\mathbf{H}$. Finally, for the lowest magnetic fields the magnetic moment of (III,Mn)V matrix saturates along the field whereas the magnetic moment of NiFe assumes an equilibrium between the effects of $\mathbf{H}$ and the shape anisotropy of the dots.

\section{CONCLUSIONS}
The micromagnetic modelling (based on the finite elements method) was applied to the films of (Ga,Mn)As ferromagnetic semiconductor with and without anti-dot structuring, as well as to (Ga,Mn)As films with NiFe inclusions. The hysteresis loops simulated for various orientations of the external magnetic field revealed the effect of anti-dot structuring on the observed in-plane magnetic anisotropy. The direction along the major axis of the elliptic anti-dots appeared to be easier (i.e. with broader hysteresis loop and larger coercive field) than the direction along the minor axis. The difference between these two directions --- measured as the ratio of demagnetization fields --- was shown to depend monotonically on the compression factor of the anti-dots. Yet the observed effect was very weak thus the modelling of (Ga,Mn)As film with the elliptic holes filled with NiFe was also performed. In such a case the structuring-induced contribution to the in-plane magnetic anisotropy was much stronger. The hysteresis loops collected along the major and minor axes of NiFe dots were qualitatively different, but in both cases revealing the governing effect of NiFe dots on the behavior of the magnetic moment of (Ga,Mn)As.

\begin{acknowledgments}
This work was supported by the European Union within  the  European  Regional  Development  Fund through the Innovative Economy program under the grant Homing Plus/2011-4/6 from the Foundation for Polish Science.
\end{acknowledgments}

\bibliography{unsrt}

\end{document}